\documentclass[a4paper,11pt]{article}
\usepackage{jcappub}

\usepackage[latin1]{inputenc}
\usepackage{amsmath}
\usepackage{amsfonts}
\usepackage{hyperref}
\hypersetup{colorlinks=True,citecolor=blue}
\usepackage{amssymb}
\usepackage{graphicx}
\usepackage{subfigure}
\usepackage[usenames,dvipsnames,svgnames,table]{xcolor}

\title{Direction dependence of cosmological parameters due to  cosmic hemispherical  asymmetry}
\author[]{Suvodip Mukherjee,}
\author[]{Pavan K. Aluri,}
\author[]{Santanu Das,}
\author[]{Shabbir Shaikh}
\author[]{and Tarun Souradeep}
\affiliation[]{Inter University Centre for Astronomy and Astrophysics \\ Post Bag 4, Ganeshkhind, Pune-411007, India}
\emailAdd{suvodip@iucaa.in}
\emailAdd{aluri@iucaa.in}
\emailAdd{santanud@iucaa.ernet.in}
\emailAdd{shabbir@iucaa.in}
\emailAdd{tarun@iucaa.in}

\abstract{\noindent Persistent evidence for a cosmic hemispherical asymmetry in the temperature field of cosmic microwave background (CMB) as observed by both WMAP as well as PLANCK increases the possibility of its cosmological origin.
Presence of this signal may lead to different values for the standard model cosmological parameters in different directions, and that can have significant implications for other studies where they are used. We investigate the effect of this cosmic hemispherical asymmetry on cosmological parameters using non-isotropic Gaussian random simulations injected with both scale dependent and scale independent modulation strengths.
Our analysis shows that $A_s$ and $n_s$ are the most susceptible parameters to acquire position dependence across the sky for the kind of isotropy breaking phenomena under study. As expected, we find maximum variation arises for the case of scale independent modulation of CMB anisotropies. We find that scale dependent modulation profile as seen in PLANCK data could lead to only $1.25\sigma$ deviation in $A_s$ in comparison to its estimate from isotropic CMB sky.}

\begin{document}

\maketitle
\section{Introduction}
Unprecedented quality of measurements of cosmic microwave background (CMB) temperature and polarization fields by several missions in the last few decades have significantly improved our understanding of the universe. The measured temperature anisotropy by WMAP \cite{wmap_param} and PLANCK \cite{planck13cl,planck13cosmopar,planck15cl,planck15cosmopar} is well explained by the minimal LCDM model. However, these missions also revealed an unforeseen signal that violates one of the fundamental pillars of cosmology viz., statistical isotropy (SI) of the cosmos.This isotropy violating signal in the CMB temperature field, known as cosmic hemispherical asymmetry (CHA), was detected at about  $3\sigma$ significance in the clean CMB maps from WMAP \cite{eriksen07,hoftuft} as well as PLANCK \cite{planck13IS,planck15IS} data. It implies that the hemisphere centered at the direction, $ \hat{p} =(l,b) = (228^\circ,-18^\circ)$ in Galactic coordinates \cite{planck13IS,planck15IS}, possess enhanced temperature fluctuations, and correspondingly suppressed fluctuations in the opposite hemisphere. Several other analysis using WMAP and PLANCK data have also confirmed this effect \cite{hansen,eriksen04,paci,akrami,quartin}. The detection of this signal in both WMAP and PLANCK data reduces its chance of arising from an unknown systematic. As a result, it is important to understand its origin, and its implication to other observational windows of cosmology. Several models have been proposed in order to incorporate this effect \cite{erickcek08a,Donoghue,dai,lyth,donald,abolhasani,namjoo,jazayeri,chluba,mukherjee15a,kothari15a,kothari15b,mukherjee15b,kenton,ashoorioon,adhikari,cai},
and many of them also predict the likely signatures perceived in polarization data as a consequence.
 
The cosmological origin of this signal will affect the cosmological parameters and can lead to direction dependence in the derived parameters. As a result, the cosmological parameters obtained from different partial sky missions may contradict each other, and can be a source of tension between different data sets.
Hence it is important to study the implications of this SI violation phenomena on cosmological parameters. Recently, an interesting study of this kind was carried out by Axelsson et al. \cite{axelsson} using WMAP $9$ year data. However unknown systematics from foreground contaminations, masking or instrumental beam can conceal the complete effects of CHA. As a result, it is important to investigate the extent to which cosmological parameters are susceptible to the change due to the cosmological origin of this anomaly from ideal simulations. The cosmological parameters estimated from a SI and the corresponding statistically non-isotropic (nSI) realization produced from the same initial seed of the Gaussian distribution helps to estimate the variation of cosmological parameters with direction, and can capture the complete effect of CHA. Effects of different scale dependencies of dipole power asymmetry can also be investigated from our analysis.  

We study the implication of CHA on cosmological parameters using ideal nSI simulations, which are free of foreground contamination, instrumental noise or asymmetric beam. An estimation of cosmological parameters from these simulations in different directions of the sky can shed light on the maximum possible variation that can be expected in these cosmological parameters due to CHA. These results can be compared with any future analysis of direction dependent cosmological parameters to carefully infer such a dependence in them, which might otherwise be possibly connected to systematics.

This paper is organized as follows. In section~\ref{bips}, we briefly review the Bipolar Spherical Harmonic (BipoSH) framework, a well suited language to quantify SI violation, that we use in this analysis. In section~\ref{sims}, we describe the procedure used to generate the simulations for our study. Results on estimation of cosmological parameters from different non-overlapping patches of the sky and conclusions are presented in section~\ref{cosmo} and \ref{conc} respectively.

\section{ BipoSH representation of SI violation}\label{bips}
The temperature and polarization field of CMB can be expressed as 
\begin{equation}\label{eqt}
X(\hat n ) = \sum_{lm} X_{lm}Y_{lm}(\hat n); \,\,\text{$X$ = T, E, B},
\end{equation}
where, $Y_{lm}(\hat n)$ are the Spherical Harmonics (SH) along the direction $\hat n$. The two point correlation function $\langle X_{lm} X^{'*}_{l'm'}\rangle$ of temperature or polarization field can be expanded in the tensor product basis of two SH spaces as
\begin{equation}\label{eq2a}
K_{lml'm'} \equiv  \bigg\langle X_{lm} X'^{*}_{l'm'}\bigg\rangle
           =  \sum_{JN} A^{JN}_{ll'|XX'}(-1)^{m'} C^{JN}_{lml-m'},
\end{equation}
where $A^{JN}_{ll'}$ are the BipoSH coefficients introduced in CMB analysis by Hajian \& Souradeep \cite{hs,bhs}, and $C^{JN}_{lml'm'}$ are the Clebsch-Gordan coefficients. Under the assumption of statistical isotropy, the covariance matrix is diagonal i.e., the only non-vanishing coefficients are $A^{00}_{ll} = (-1)^l\sqrt{2l + 1}C_l$ corresponding to $J=0$, $N=0$, where $C_l$ is the familiar angular power spectrum. For the case of nSI CMB sky, off-diagonal terms of the covariance matrix are also non-zero, which in turn lead to non-zero BipoSH coefficients for $J>0$.

\begin{figure}
\centering
\includegraphics[width=4.0in,keepaspectratio=true]{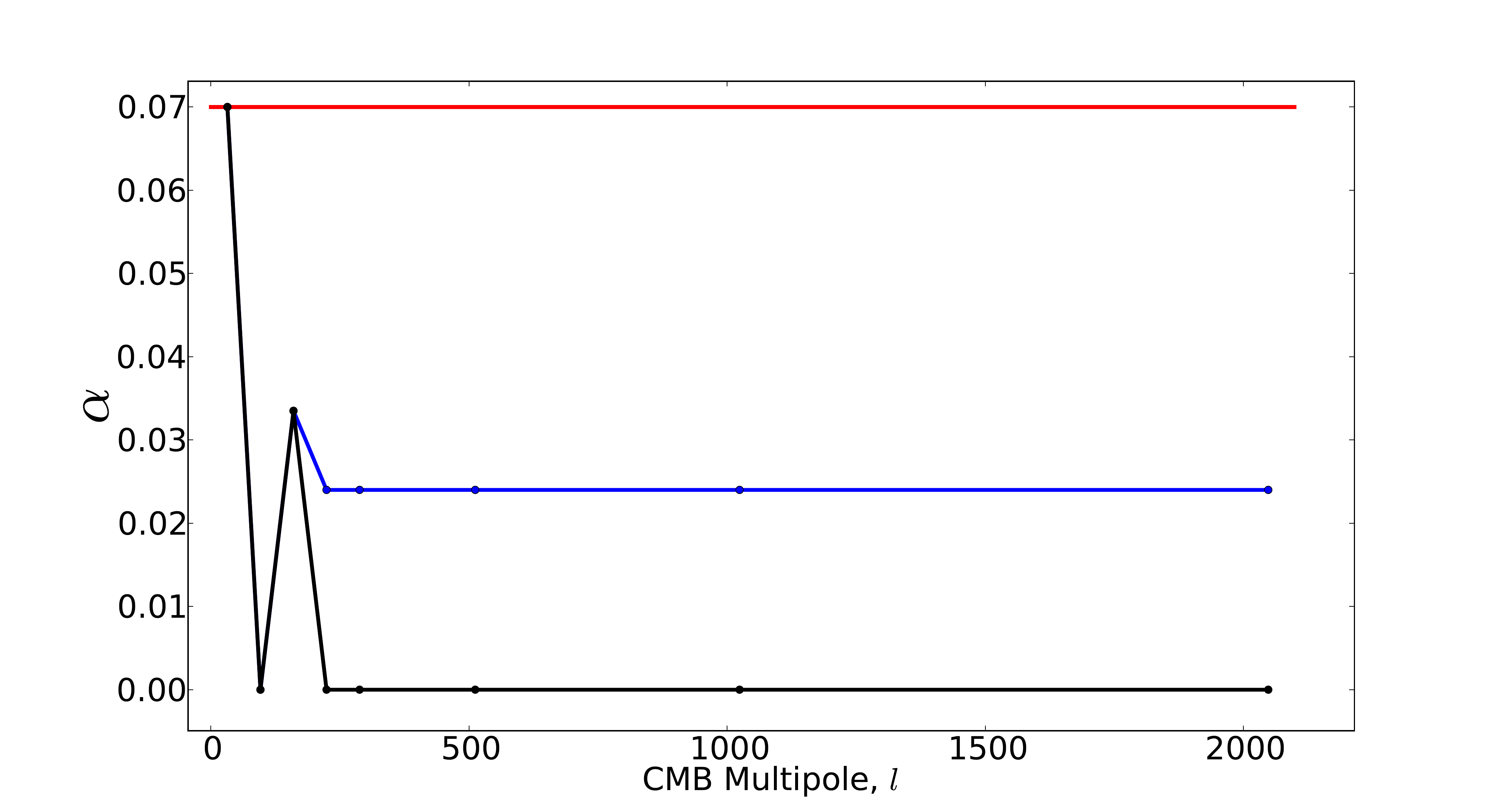}
\caption{The profiles of amplitude of modulation for the three different cases
		studied in the present work -
        $(i)$ SD-1 : a scale dependent modulation profile with some residual anisotropy at
        small scales, $(ii)$ SD-2 : scale dependent profile as seen in PLANCK data, and
        $(iii)$ SID : a scale independent modulation, are shown here in blue, black and
        red respectively.}
\label{fig1}
\end{figure}

The observed CHA can be modelled as \cite{gordon, gordon-t}
\begin{equation}\label{eq2b}
\tilde T(\hat n)= T(\hat n)(1 + \alpha \, \hat{p} \cdot \hat{n}),
\end{equation}
where $T(\hat n)$ and $\tilde T(\hat n)$ are the unmodulated and modulated CMB anisotropies in the sky direction $\hat{n}$. $\alpha$ is the modulation strength and $\hat p$ denotes the direction of modulation field. In our analysis, we also consider the modulation field to be scale dependent $\alpha \equiv \alpha_l$ to mimic the observation by WMAP \cite{eriksen07,hoftuft} and PLANCK \cite{planck13IS,planck15IS}. This model generates only $J=1$  BipoSH coefficients given as \cite{planck13IS, mukherjee15a}
\begin{equation}\label{eqbi}
A^{1N}_{ll+1 |TT} = \mathcal{M}^{1N}[C^{TT}_l + \, C^{TT}_{l+1}]\frac{\Pi_{ll+1}}{\sqrt{4\pi}\Pi_{1}}C^{10}_{l\,0l+1\,0},
\end{equation}
where $\mathcal{M}^{1N}$ are the SH coefficients of the modulation field $\mathcal{M}(\hat n) = \alpha\, \hat p\cdot \hat n$, $\Pi_{l_1l_2\dots } = \sqrt{(2l_1+1)(2l_2+1)\dots}$ and $C^{TT}_l$ is the temperature angular power spectra. The modulation amplitude $\alpha$ is related with $\mathcal{M}^{1N}$ by the relation $\alpha = 1.5\sqrt{\frac{\sum_{N} |\mathcal{M}^{1N}|^2}{3\pi}}$. In Fig.~\ref{fig1} we show, three kinds of modulation profiles considered in the present work viz., ($i$) a scale dependent (SD-1) modulation strength depicted by blue curve with some residual power at high $l$, ($ii$) a scale dependent (SD-2) modulation strength depicted by black curve to mimic the observed profile in PLANCK data \cite{planck13IS,planck15IS}, and ($iii$) a scale independent (SID) modulation depicted by red curve in the same figure. Using the BipoSH coefficients for these two different cases, we produce nSI Gaussian simulations as described in the next section.

\section{Simulations}\label{sims}
To study the effect of dipole power asymmetry on cosmological parameters, we produce dipole modulated CMB skies at \texttt{HEALPix}\footnote{\url{http://healpix.jpl.nasa.gov/}} \cite{healpix} resolution parameter of $N_{side}=1024$ using \texttt{CoNIGS} (Code for Non-Isotropic Gaussian Sky). \texttt{CoNIGS} is a numerical algorithm developed by Mukherjee \& Souradeep \cite{conigs} to produce statistically anisotropic Gaussian realizations of CMB.
This algorithm implements Cholesky decomposition of the covariance matrix, $K_{lml'm'}$ in Eq.~\ref{eq2a} which contains both angular power spectra $C_l$ and BipoSH coefficients $A^{1N}_{ll'}$, and obtains the corresponding lower triangular matrix. Operating this lower triangular matrix on an array of random draws from a unit variance Gaussian distribution produces the simulations which have inherently broken statistical isotropy with given BipoSH spectra as ensemble mean. With input as angular power spectra (consistent with PLANCK) and BipoSH coefficients due to CHA, we generate nSI Gaussian simulations for our analysis. 

\begin{figure}
\centering
\includegraphics[width=4.0in,keepaspectratio=true]{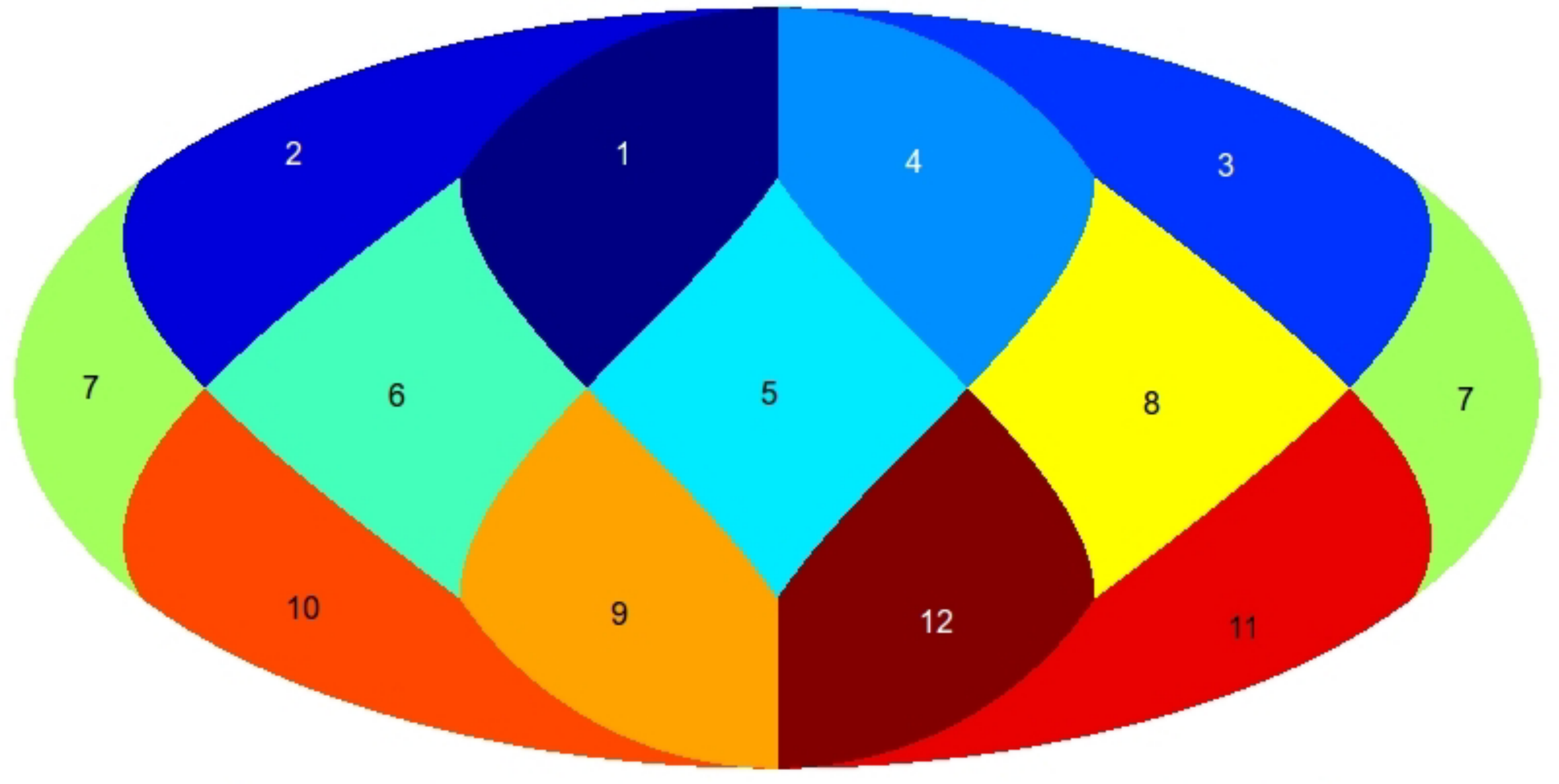}
\caption{Full sky realizations of CMB are partitioned into $12$ different regions as
         depicted in this diagram. Cosmological parameters are estimated from these
         $12$ different patches from SI and nSI simulations. The anisotropic simulations
         are produced with cosmic hemispherical asymmetry injected in the direction of
         pixel center of patch~$12$.}
\label{fig2}
\end{figure}

Simulations with both scale dependent and scale independent modulation amplitudes as shown in Fig.~\ref{fig1}, are generated. These simulations are free from any foreground contamination, instrumental noise and asymmetric beam. Hence this analysis is not affected by any systematics.  The corresponding isotropic simulations are produced with the same seed and theoretical $C_l$, used to produce nSI simulations.  To study the effect of statistical anisotropy on cosmological parameters, we need to extract the angular power spectrum, $C_l$, from different sky directions. We divide the sky into mutually exclusive partitions
as shown in Fig.~\ref{fig2}.

\begin{figure}[h]
\centering
\subfigure[]{
\includegraphics[width=3.0in,keepaspectratio=true]{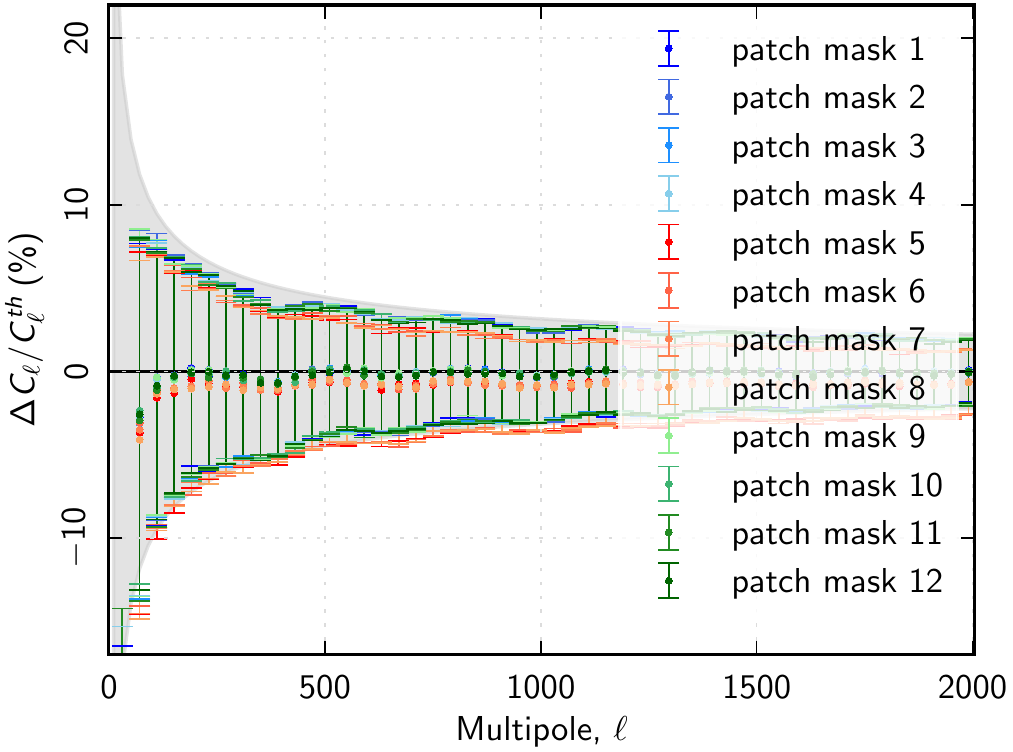}\label{fig3}
}
\subfigure[]{
\centering
\includegraphics[width=3.0in,keepaspectratio=true]{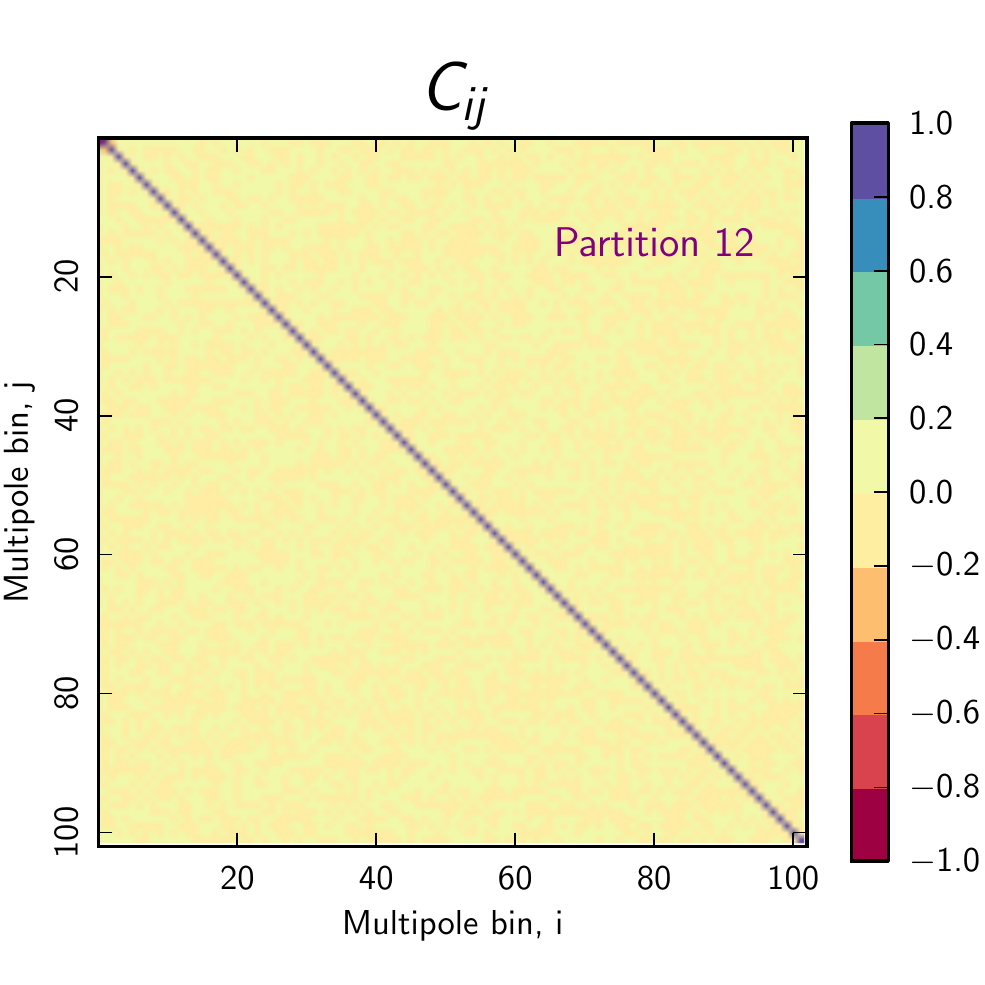}\label{fig3a}
}
\caption{(a)~Percent difference plot of the recovered mean $C_l$ with error bars from different patches of 1000 SI simulations in comparison to the input theoretical $C_l$ used to generate the simulations. The convergence to the injected $C_l$ improves with $l$ and are well within the cosmic variance $=\sqrt{2/(2l+1)}\times 100 (\%)$.
(b)~The covariance matrix of binned $C_l$ obtained from $1000$ simulations from mask region~12 is shown here. Bin index `$i$' contains the multipoles $[(i-1)*\Delta l+2,i*\Delta l+1]$, where $\Delta l=20$ is the chosen bin width. One can clearly see that the covariance matrix is sufficiently diagonal.}\label{fig3b}
\end{figure}

To assess the maximum effect induced by isotropy violation on cosmological parameters, we inject the CHA in the direction of one of the pixel centers of \texttt{HEALPix} $N_{side}=1$ grid, which are our sky partitions shown in Fig.~\ref{fig2}.
Specifically we choose pixel center corresponding to patch~$12$ as the direction for injecting dipole modulation in our simulations. The pseudo$-C_l$ MASTER algorithm \cite{master} is used to recover angular power spectrum from each of these patches in the multipole range $l=[2,2048]$ using a bin width of $\Delta l = 20$. Due to the limited size of these sky patches, the recovered $C_l$ are more accurate at high `$l$' as shown in Fig.~\ref{fig3}. To avoid artefacts due to sharp boundaries of the sky partitions in $C_l$ recovery, we anodise each partition shown in Fig.~\ref{fig2} with a Gaussian beam of FWHM=50' (arcmin) and requiring $\beta \sim 10^{-3}$
following \cite{Kim2011}. The estimated $C_l$ from the central patches ($5,\,6,\,7,\,8$) shows a mild under estimate of power. However, the recovered $C_l$ are completely consistent within the cosmic variance as shown in Fig.~\ref{fig3}.
The covariance matrix of the binned $C_l$ obtained from $1000$ SI simulations with a bin width of $\Delta l = 20$, from the patch ~12, is shown in Fig.~\ref{fig3a}.
It is defined as $C_{ij} = \langle \delta C_{l_i} \delta C_{l_j} \rangle/\sqrt{\langle \delta C_{l_i}^2 \rangle \langle \delta C_{l_j}^2 \rangle} $, where $\delta C_{l_i} = C_{l_i} - \langle C_{l_i} \rangle$ and $C_{l_i}$ is the recovered angular power spectrum from partial sky from the multipole bin `$i$'. The range of multipoles that correspond to bin $i$ are $[(i-1)*\Delta l+2,i*\Delta l+1]$, and $l_i$ denotes the central multipole of this bin. We also checked the covariance matrices of the recovered $C_l$ from other sky partitions, and find them to be predominantly diagonal.	

Due to SI violating dipole modulation, $C_l$ recovered from different patches would differ for SI and nSI simulations, which in turn affect the cosmological parameters from them.
In our analysis, we don't consider $C_l$ recovered at low multipoles $l=[2,41]$ for estimating the cosmological parameters estimation for SID and SD-1 case. For SD-2 case the effect of modulation is present only at low $l$, so we consider the full range of $l=[2,2048]$ in our analysis.
In the next section, we discuss the results for the four different cases.

\section{Assessing directional dependence of cosmological parameters}\label{cosmo}
From the angular power spectra obtained from the $12$ sky partitions, we estimate cosmological parameters for all the four cases : SI (statistical isotropy), SD-1 (scale dependent with some constant power at high $l$), SD-2 (scale dependent with no power at high $l$) and SID (Scale independent). As these ideal simulations are free from any instrumental noise or systematics, the variance of angular power spectra is solely due to cosmic variance. As a result, we can define the likelihood as
\begin{equation}
-\log \mathcal{L}(C_l|\bar C_l')= \sum_{ll'}(C_l-\bar C_l')G_{ll'}^{-1}(C_l-\bar C_l')^T\,,
\end{equation}
where, $G_{ll'}$ denotes the signal cosmic covariance matrix for $C_l$. In our case, $G_{ll'}$ is a diagonal matrix containing only cosmic variance. $C_l$ indicates the model power spectra while $\bar{C}_l$ indicates
the recovered power spectrum from different patches.
\begin{figure}[h]
\centering
\subfigure[]{
\includegraphics[width=2.5in,keepaspectratio=true]{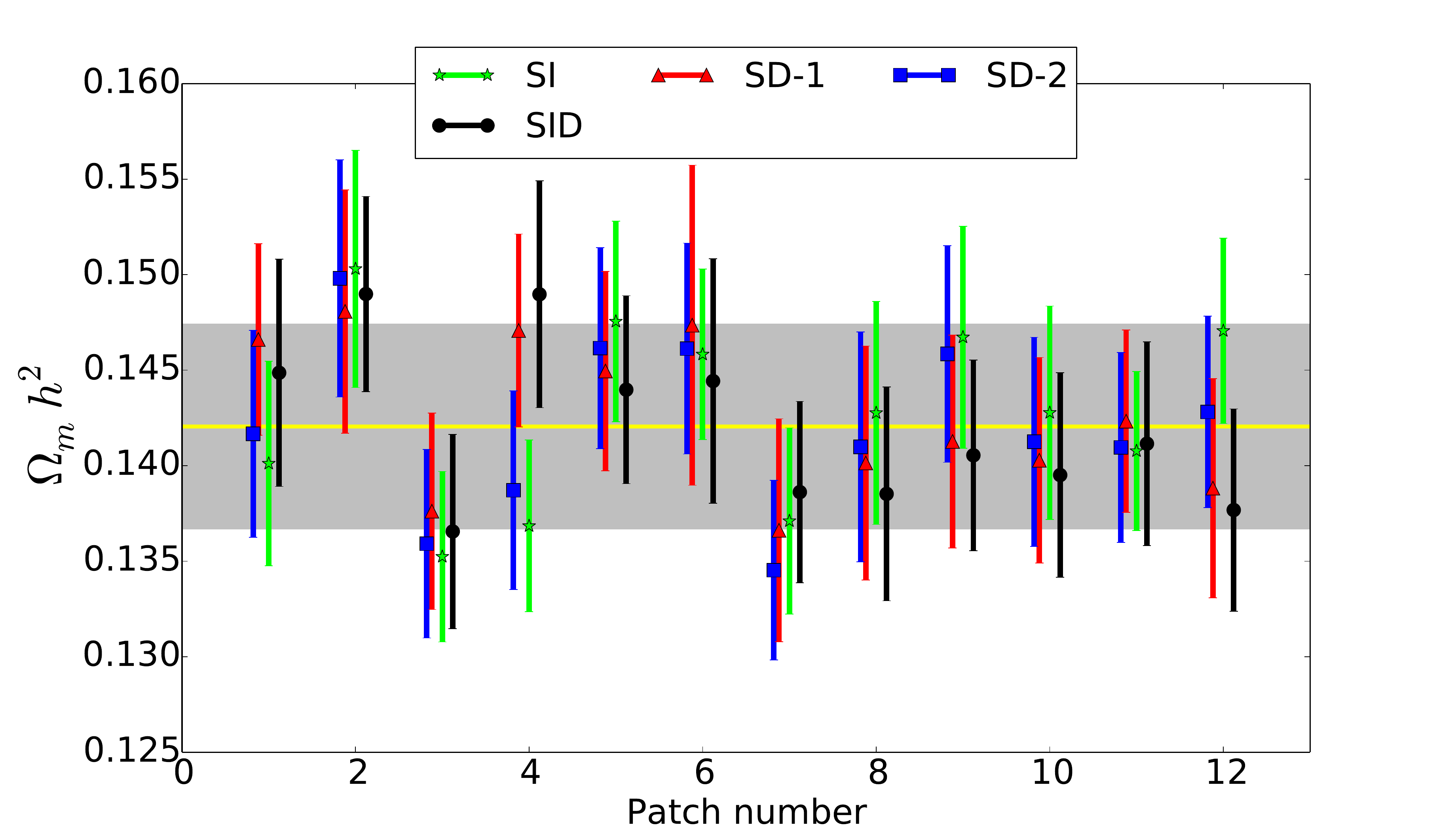}\label{figom}
}
\subfigure[]{
\includegraphics[width=2.5in,keepaspectratio=true]{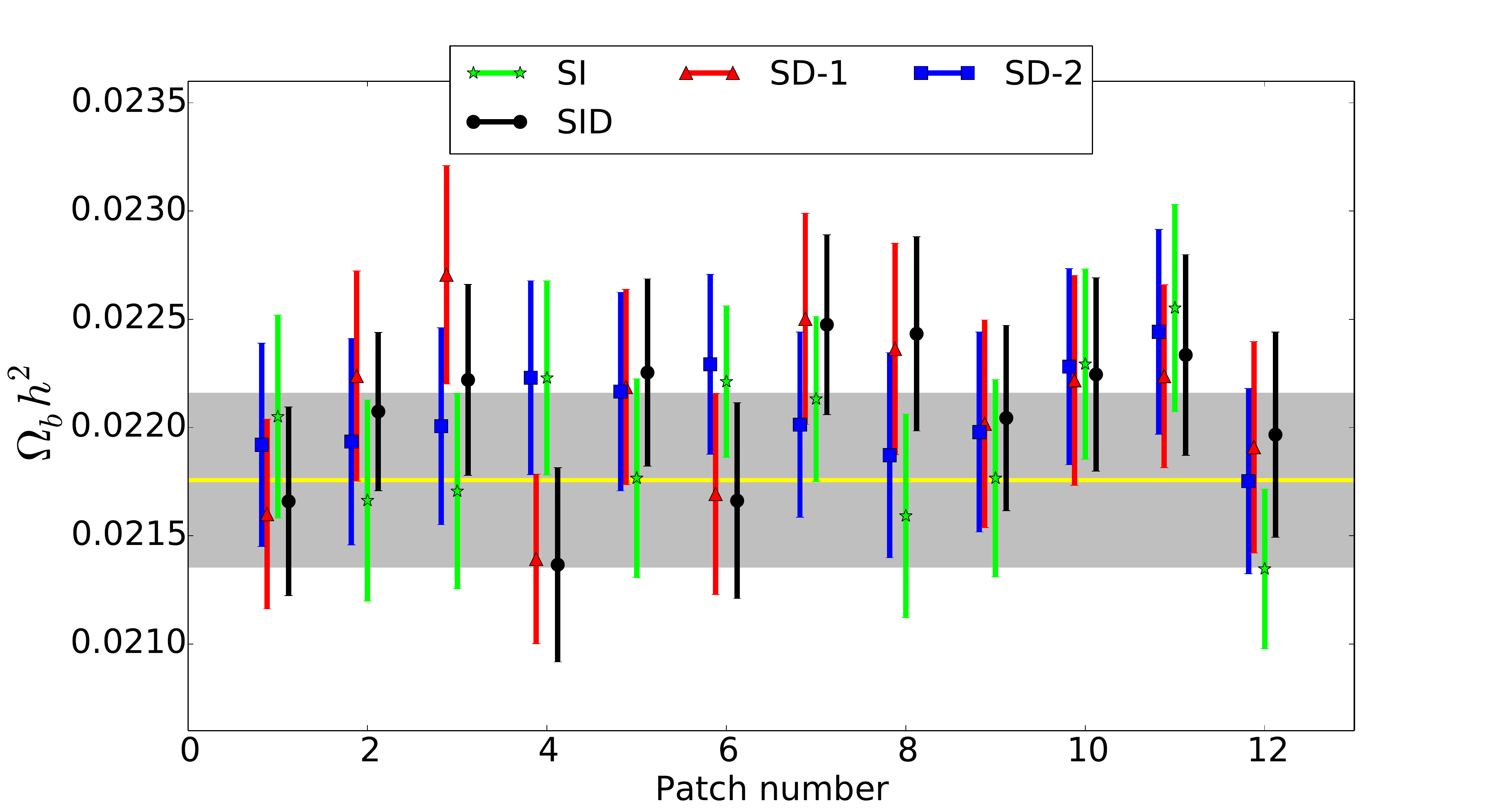}\label{figob}
}
\subfigure[]{
\includegraphics[width=2.5in,keepaspectratio=true]{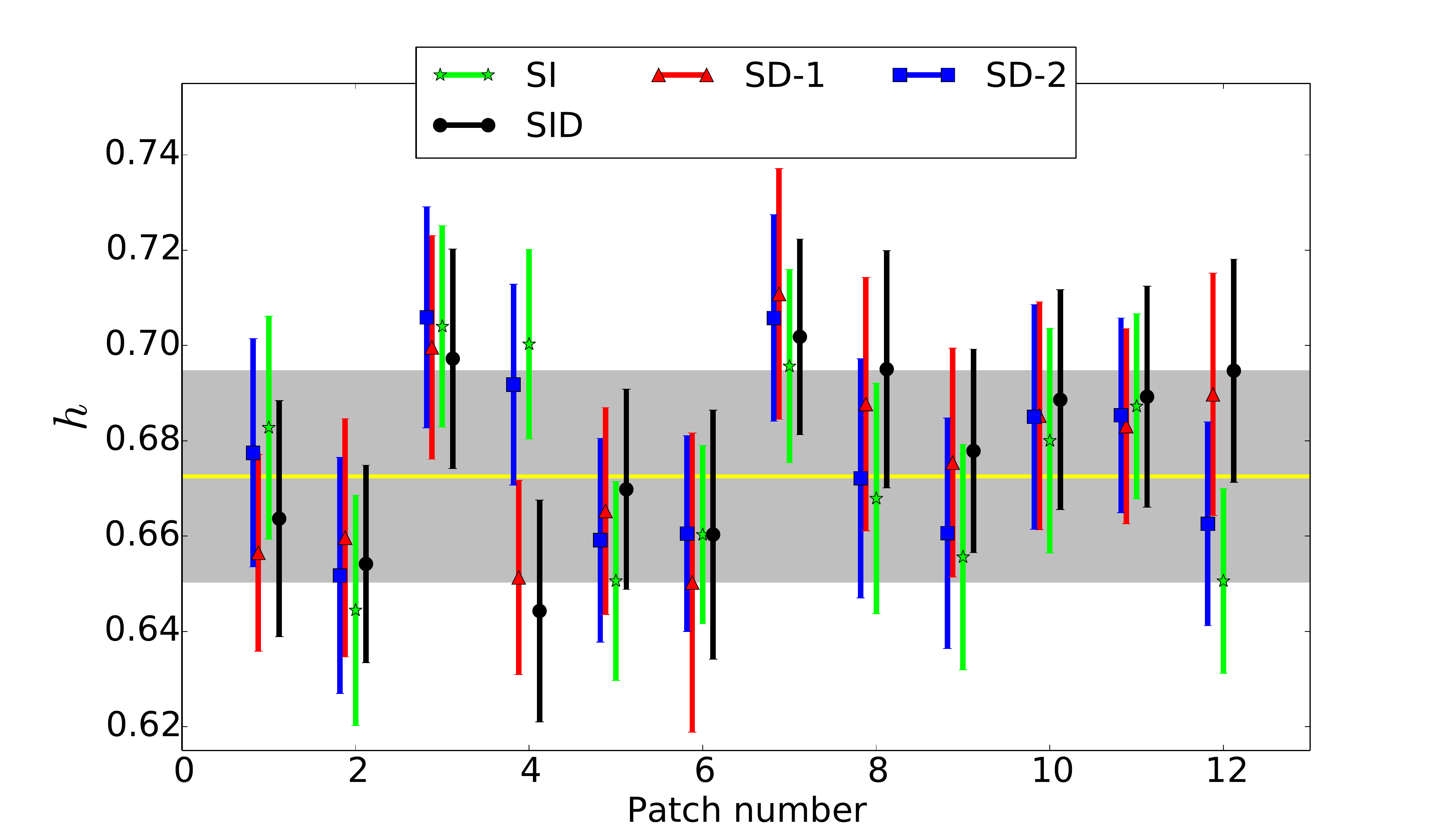}\label{figh}
}
\subfigure[]{
\includegraphics[width=2.5in,keepaspectratio=true]{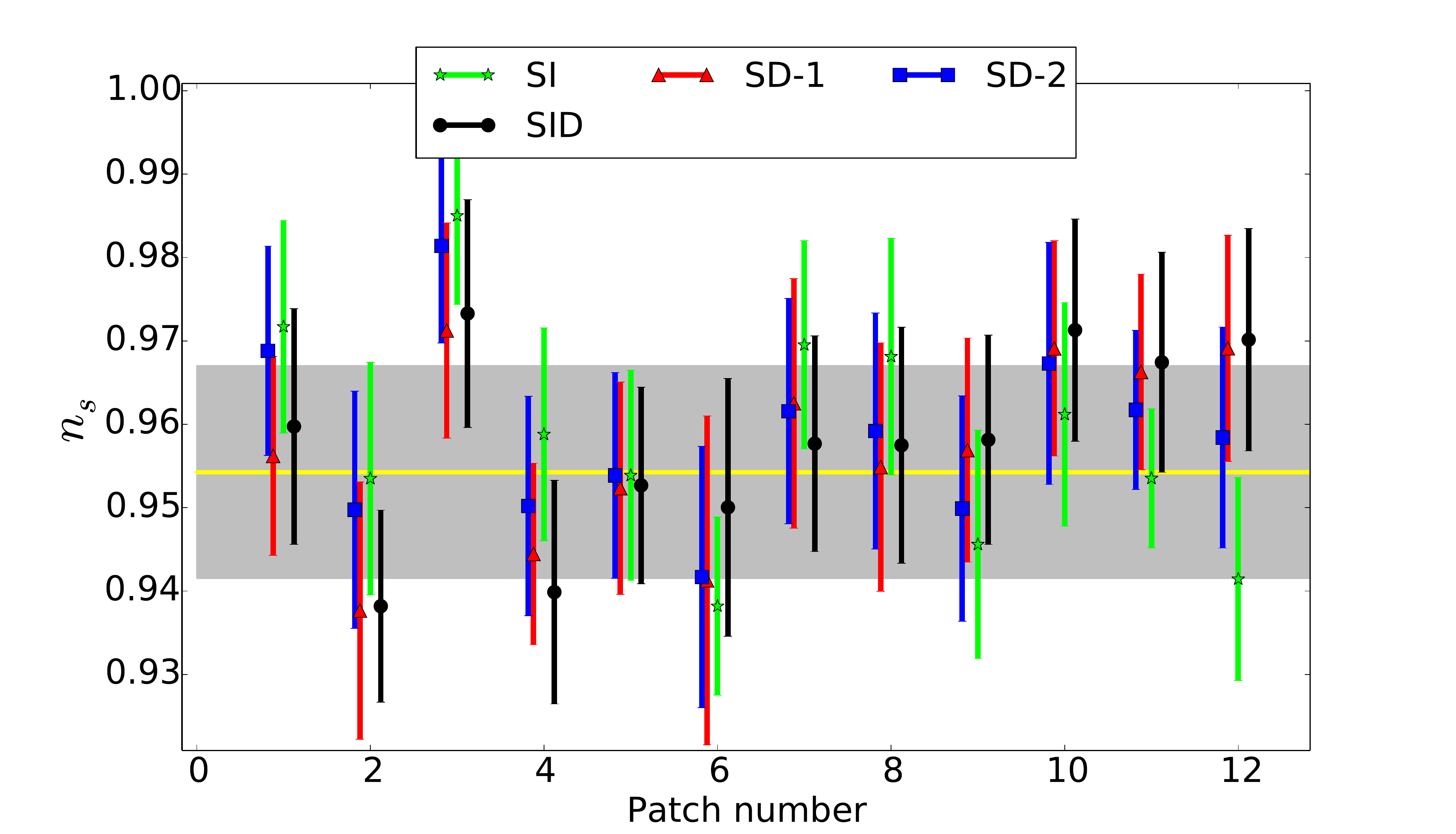}\label{figns}
}
\subfigure[]{
\includegraphics[width=2.5in,keepaspectratio=true]{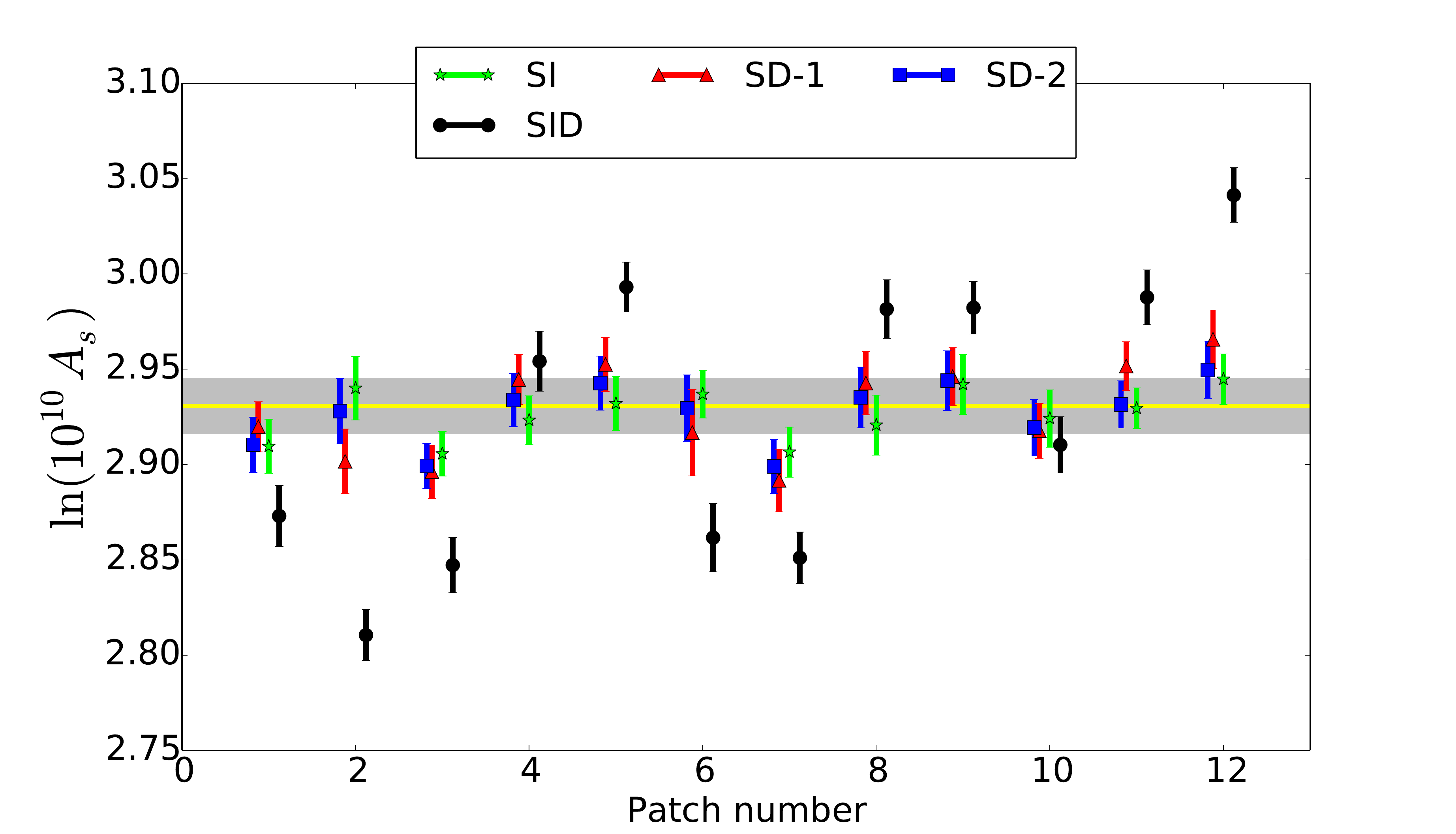}\label{figas}
}
\caption{The mean and standard deviation of the best-fit cosmological parameters $\{\Omega_mh^2, \Omega_bh^2, h, n_s, A_s \}$ ($\tau$ is excluded) obtained from the $12$ sky partitions (shown in Fig.~\ref{fig2}) of a single realization that is $(i)$ Statistically Isotropic (SI) (green), $(ii)$ anisotropic with scale independent (SID) modulation profile (black), $(iii)$ anisotropic with SD-1 profile (red), and $(iv)$ anisotropic with SD-2 profile (SID, SD-1 and SD-2 dipole modulation profiles are shown in Fig.~\ref{fig1}). In grey, we show the $1\sigma$ region for each parameter estimated from the full sky SI simulation with the mean depicted in yellow. Same seed is used to generate all the four CMB sky maps viz., SI, SID, SD-1, SD-2 to allow direct comparison of the derived cosmological parameters. None of the parameters, except for the scalar amplitude $A_s$, show any significant variation with sky direction, in comparison to that arising due to cosmic variance in a statistically isotropic model.}
\label{figparam}
\end{figure}

From the nSI maps produced with SD-1, SD-2 and SID modulation profiles, we obtain the best-fit standard model cosmological parameters $\{\Omega_mh^2, \Omega_bh^2, h, \tau, n_s, A_s\}$, from all partitions of the sky using \texttt{SCoPE} (Slick Cosmological Parameter Estimator), developed by Das \& Souradeep \cite{scope}. We also obtain the same set of six cosmological parameters from all the $12$ patches from the corresponding SI simulation with the same seed. The parameter estimation uses $C_l$ in the range $l=[40,2048]$ for SI, SD-1 and SID case and $l=[2,2048]$ for SD-2 case. Since we limit to only temperature data, $\tau$ cannot be constrained by this analysis.
\begin{table*}[t]
\centering
\caption{Summary on departure of cosmological parameters due to dipole power asymmetry
         (in units of $\sigma$) as seen in $n$SI case compared to full sky SI case.}
\label{table1} 
\vspace{0.5cm}
\begin{tabular}{|p{1.5cm}|p{2.5cm}|p{2.5cm}|p{2.5cm}|p{4.50cm}|}
\hline 
 \centering Parameter &  \centering Maximum relative departure seen in SD-1 case &  \centering Maximum relative departure seen in SD-2 case & \centering   Maximum relative departure seen in SID case & \centering Remarks
\tabularnewline
\hline
$A_s$ & $2.27\sigma$ for patch $12$ and $1.7\sigma$ for patch $2$ & $1.25\sigma$ for patch $12$ and $0.16\sigma$ for patch $2$ & $7.71\sigma$ for patch $12$ and $8.93\sigma$ for patch $2$ & \centering Important for SID. Not significant for SD-1 and SD-2.  \tabularnewline
\hline
$n_s$ & $ 1.1\sigma$ for patch $12$ & $ 0.31\sigma$ for patch $12$ & $1.19\sigma$ for patch $12$ & \centering Not significant. \\ Indicates a mild increase (or decrease) in the direction (or opposite) of cosmic hemispherical asymmetry. \tabularnewline
\hline
$h$ & $1.04\sigma$ for patch $4$ &  $0.9\sigma$ for patch $4$ & $1.21\sigma$ for patch $4$ & \centering Not significant \tabularnewline
\hline
$\Omega_mh^2$ &  $0. 93 \sigma$ for patch $4$ & $1.06\sigma$ for patch $4$ & $0.87\sigma$ for patch $4$ & \centering  Not significant   \tabularnewline
\hline
$\Omega_bh^2$ & $ 0.94 \sigma$ for patch $2$ & $ 1.23 \sigma$ for patch $2$ & $1.36\sigma$ for patch $2$& \centering  Not significant \tabularnewline
\hline
\end{tabular}
\end{table*}

A comparison of the estimated cosmological parameters from various patches are depicted in Fig.~\ref{figparam} for SI (green), SD-1 (red), SD-2 (blue) and SID (black) cases from a single CMB realization with same seed.
The use of same seed for generating an SI map as well as nSI maps with SID, SD-1 and SD-2 modulation profiles, allows us to understand the directional dependence in the inferred cosmological parameters corresponding to each of these SI violation scenarios.

In this figure, we plotted the mean along with $1\sigma$ variation of all the five parameters, derived from posteriors constructed using \texttt{SCoPE} chains, as a function of  sky partition number. As can be seen from Fig.~\ref{figparam}, the maximum discrepancy is observed for the parameter $A_s$.
A large positive departure from the SI simulation estimate is evident only in patch~$12$, for SID case, as this is the direction of the injected dipole power asymmetry. Correspondingly a large negative shift for $A_s$ is seen in the direction of patch~$2$. All other patches adjacent to patch numbers $12$ and $2$ also show a drift towards higher and lower value of $A_s$, respectively. Estimates from patch number $10$ and $4$ do
not show any notable departure. On comparison, for a $7\%$ amplitude of dipole modulation in SD-1 and SD-2 cases at low~$l$, the recovered parameters are found to be completely consistent with SI map within $1.5\sigma$. SID case shows a severe departure as expected. But PLANCK-2015 results \cite{planck15IS} clearly indicate that, the amplitude of observed dipole modulation falls off beyond $l \sim 64$.

In Table~\ref{table1}, we list the discrepancy between cosmological parameters derived from SID, SD-1 and SD-2 cases, individually in comparison to the parameters estimated from full sky SI map (except for $\tau$ which cannot be constrained using only temperature data).
It is also important to note that, although the departure of $n_s$ in any of the patches from SI case is at best $\sim 1.2\sigma$, there is a clear trend of excess/deficit in the same/opposite direction of dipole power asymmetry. In the hemisphere with more fluctuations, $n_s$ indicates a mild increase for the SID modulation case, whereas in the opposite hemisphere there is a reduction in its value. This is opposite to the trend observed by Axelsson et al. \cite{axelsson} in the WMAP data. If a similar behaviour is observed for $n_s$ in PLANCK data, then it may have an origin different from CHA.
It may well be due to foreground contamination. Nevertheless, these estimates of $n_s$ from different patches/sky directions for both SD and SID cases are in good agreement with those derived from full sky SI map.

To understand the statistical nature of the nSI signal, we estimate the excess (deficit) power in the direction $\hat{p}$, i.e., towards the center of patch 12 (in $-\hat{p}$ direction which corresponds to center of patch 2) from $1000$ simulations using local variance estimator proposed by Akrami et al. \cite{akrami}.
Histogram of recovered dipole amplitudes from the normalized variance variation maps ($\xi (\hat{N})$) in both directions, obtained from each of the 1000 nSI maps corresponding to SD-2 and SID cases are plotted in figure Fig.~\ref{figdis1} and Fig.~\ref{figdis2}), respectively.
A brief description of this procedure is given in Appendix~\ref{app1}. From Fig.~\ref{figdis2}, we see that the mean of the recovered dipole amplitudes in $\hat{p}$ and $-\hat{p}$ directions of variance variation maps thus obtained from simulations match with the injected amplitude $2A=0.14$. The histogram also indicates that there are many simulations with much lower dipole amplitude than $0.14$. Hence the estimated parameters, particularly $A_s$, from a given map/realization can be smaller than $14\%$ in the direction of anisotropy as was found here. In the SID case, we find only a $4\%$ excess (deficit) in $A_s$ in patch 12 (patch 2) which is within $1\sigma$ of the mean dipole amplitude obtained from 1000 simulations as shown in Fig.~\ref{figdis2}. A similar effect is also observed in the SD-1 case which has a residual dipole anisotropy of $\sim 2.5\%$ in amplitude at high~$l$. We find that the estimated value of $A_s$ from patch 12 from the specific realization we used is also lower (than $5\%$), but within $1\sigma$ of the dipole amplitude distribution shown in Fig.~\ref{figdis1}. Thus we can understand this recovery of lower value for $A_s$ than the expected/injected value along the axis of dipole modulation (i.e., from sky partitions 12 and 2) as a manifestation of the statistical nature of the underlying anisotropic signal.

Using WMAP data, Axelsson et al. \cite{axelsson} also found a mild direction dependence for $\Omega_bh^2$. But our analysis on ideal simulations that compares nSI map with corresponding SI map does not show any such departure and hence does not have any discernible pattern for $\Omega_bh^2$ due to CHA.
\begin{figure}
\centering
\subfigure[]{
\includegraphics[width=0.45\textwidth]{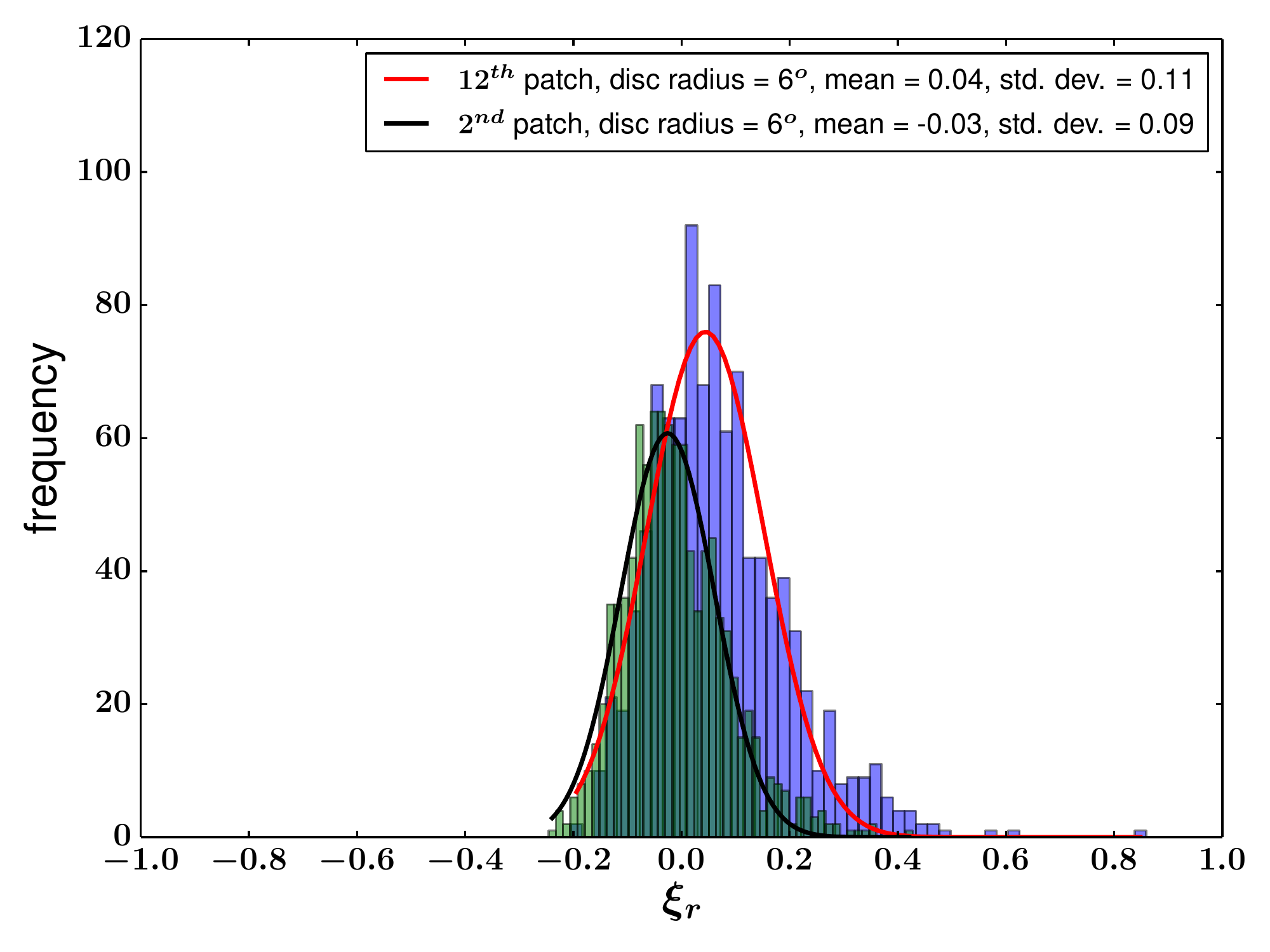}\label{figdis1}
}
\subfigure[]{
\includegraphics[width=0.45\textwidth]{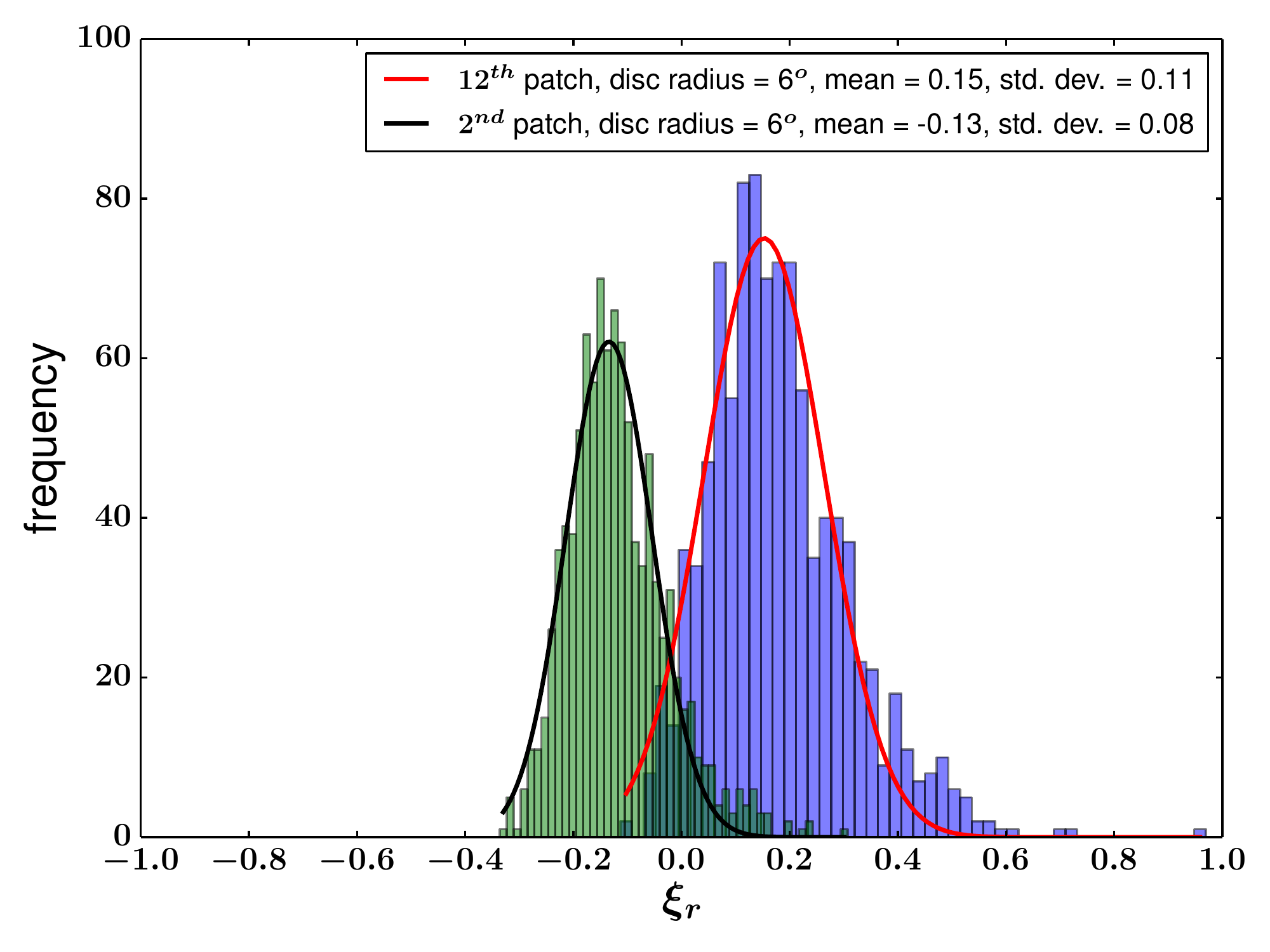}\label{figdis2}
}
\caption{Histogram of dipole amplitudes recovered from $\hat{p}$ and $-\hat{p}$ directions of the normalized variance variation maps, $\xi(\hat{N})$, obtained from 1000 nSI maps injected with (a) SD-2 and (b) SID modulation profiles.
The distributions are overlaid with a Gaussian function with the mean and standard deviation mentioned in the plot.}
\label{fignsid}
\end{figure}

Other parameters does not show any significant departure or special trend in comparison to the SI case. This is similar to the finding by \cite{axelsson} using WMAP data. All other parameters shows variations which are well within $1\sigma$ of the SI simulation and doesn't show any variation.
To summarize, the current analysis indicates that the presence of dipole power asymmetry at an amplitude of $7\%$  seem to affect the cosmological parameters related to inflationary physics i.e., $A_s$ and $n_s$. Rest of the parameters appear to be not influenced by its presence.

\section{Conclusion}\label{conc}
The estimation of cosmological parameters from noise and foreground free dipole modulated statistically non-isotropic (nSI) simulations allows an assessment of  the maximum possible departure that can arise in such a nSI map, in comparison to their estimates from a corresponding statistically isotropic (SI) sky. For several partial sky missions of CMB and other cosmological/astrophysical missions with limited sky coverage, it is essential to be aware of any departure in the parameters compared to SI model. Our analysis sheds light on this specific issue by estimating the cosmological parameters from sky partitions in different directions, and identifies the parameters most susceptible to the presence of a dipolar modulation SI violation. We perform the estimation in the $12$ different sky partitions shown in Fig.~\ref{fig2} using nSI and corresponding SI simulation. The nSI simulations are produced using \texttt{CoNIGS} algorithm with dipole modulation that is scale dependent (as observed by PLANCK), as well as, the simpler scale independent  amplitude profiles as shown in Fig.~\ref{fig1}.
The injected direction for the modulation in both the cases is taken to be the center of the patch mask~$12$ from Fig.~\ref{fig2}, to maximize the effect of SI violation and amplitude is taken to be $0.07$ consistent with that indicated in the recent PLANCK cosmological results.

The parameter that is most susceptible to variation on the sky due to SI violation is $A_s$. As expected, it clearly predicts a higher value in the direction of modulation and a lower value in the opposite direction. Variations as large as  $7.7\sigma$ and $8.93\sigma$  deviations are observed in $A_s$ from patches $12$ and $2$, respectively, for SID case. However, for the SD-2 model, which mimics the observed modulation profile in CMB maps from PLANCK observations, only a $1.25\sigma$ deviation is seen in patch $12$. This indicates that the estimates from SD-2 model would not differ significantly from those obtained from a full sky SI map. However, if the  scale dependent modulation is extended to high~$l$ as in SD-1 case, then a $2.27\sigma$ and $1.7\sigma$ departures were seen in the sky directions 12 and 2, respectively, in comparison to the SI map estimates. This is due to  the fact that the cosmological parameters are heavily influenced by $C_l$ at high~$l$, rather than low~$l$ power spectrum. The other parameter that is found to be mildly susceptible to SI violation is $n_s$. It does not show any significant departure in simulations with SD-1, SD-2 and SID modulation profiles compared to SI case. But it indicates a higher value in the direction of CHA than in the opposite direction. Rest of the parameters do not show any significant variation correlated with the presence of CHA. In essence, the inflationary parameters, $A_s$ and $n_s$, are found to be more susceptible to change with direction than other parameters, due to dipole power asymmetry.

In summary, we conclude that the base LCDM model parameters does not show any direction dependence, as long as the modulation amplitude of $7\%$ is scale dependent and is present only at large angular scales($>3^\circ$) \cite{planck13IS,planck15IS}. The determination of the true nature of CHA from future analysis \cite{mukherjeeha}, can give a better  estimate on the effect of $A_s$ and $n_s$. Our local motion through CMB rest frame ($\vec{\gamma} = \vec{v}/c$) also induces (a scale independent) SI violation \cite{kosowsky,amendola,mukherjee14} with a magnitude of $\mathcal{M} \equiv |\gamma| = 0.00123$ \cite{kogut}, that is much weaker than the modulation strength considered here.
As a result, all the cosmological parameters can be safely considered to be unaffected by Doppler boosting of CMB anisotropies. Future CMB and other cosmological surveys from diverse observational windows may reveal the true origin/extent of this observed CHA in CMB. Recently, an analysis on Lyman$-\alpha$ forest \cite{hazra} from $z>2$  quasar data set of SDSS-III BOSS DR$9$ indicates no significant ($>3\sigma$) departures from SI model. A similar study on SNIa was also carried out recently \cite{javanmardi}.
The current analysis using simulated ideal nSI maps will be useful in carefully inferring evidence for direction dependent cosmological parameters from those observations.

\textbf{Acknowledgement :}
S.~Mukherjee and S.~Das thanks Council of Scientific \& Industrial Research (CSIR), India for
financial support as Senior Research Fellows. S.~Shaikh thanks University Grants Commission (UGC),
India for providing the financial support as Senior Research Fellow. The present work is carried
out on the High Performance Computing facility at IUCAA. The authors acknowledge the use of MASTER
routine from the Cosmology Routine Library in Fortran (CRL) webpage of  E. Komatsu (\url{http://www.mpa-garching.mpg.de/~komatsu/crl/}). 

\appendix
\section{Local Variance Estimator}\label{app1}
CHA leads to asymmetry in variances computed locally on a CMB temperature map. This relation between CHA and local variances was used by Akrami et al. \cite{akrami} to obtain the amplitude $(A)$ and direction $(\hat {p})$ of the low~$l$ hemispherical asymmetry modelled as dipole modulation of CMB anisotropies. The local variance $\sigma^2_r (\hat{N})$ in the direction $\hat{N}$ refers to the variance of temperature map computed using
pixels falling within a circular disc of a chosen radius `$r$' centred at $\hat{N}$ i.e.,
\begin{equation}
\sigma^2_r (\hat{N}) = \sum_{\hat{n} \in r} (T(\hat{n}) - T_r (\hat{N}) )^2 \,,
\end{equation}
where $T_r (\hat{N})$ is the mean of all pixels inside the radius `$r$' of the circular disc defined at $\hat{N}$.
In practice, a local variance map is obtained from a temperature map at higher resolution with HEALPix resolution parameter $\texttt{nside}=N_{side, high}$, using a low resolution HEALPix grid with $\texttt{nside}=N_{side, low}$ ($< N_{side, high}$). The low resolution HEALPix grid is used to define the disc centres $\hat{N}$ at which we compute the local variances. The variances thus computed locally at $\hat{N}$ are assigned as pixel values to the corresponding HEALPix pixel indices, to obtain a local variance map at $\texttt{nside}=N_{side, low}$.

When computing local variances in real data, one has to use an effective disc  obtained by multiplying the
    circular disc of radius 'r' defined locally at the sky position $\hat{N}$ and the galactic mask so as to avoid biases
    due to foreground contamination. Since less number of pixels will be available in the net disc , one can impose
    an additional criteria of using that location in fitting the dipole signal only if at-least 50\% (for example) of pixels
    are available for computing the local variance,  after taking the union of circular disc defined at that location and
    the galactic mask, compared to the total number of pixels available in the full disc of radius 'r'. For a dipole modulated map given by Eq.~\ref{eq2b} which is otherwise isotropic, local variances are related by $\tilde{\sigma}^2 (\hat{N}) \approx (1+2\alpha \, \hat{p} \cdot \hat{N}) \sigma^2 (\hat{N})$, where $\tilde{\sigma}^2$ and $\sigma^2$ denote local variances corresponding to modulated and unmodulated CMB sky respectively (computed using a circular disc of chosen radius `$r$' centred in the sky direction $\hat{N}$).
Thus, in order to obtain a correct estimate of the dipole signal lying beneath the observed sky, we define a normalized variance variation map as
\begin{equation}
\xi_r (\hat{N}) = \frac{\tilde{\sigma}^2_r (\hat{N}) - \langle \sigma^2_r (\hat{N}) \rangle}{\langle \sigma^2_r (\hat{N}) \rangle}\,,
\end{equation}
where $\langle \sigma^2_r (\hat{N}) \rangle$ denotes mean variance map obtained from an ensemble of SI simulations.
By fitting a dipole to this normalized variance map, $\xi_{r}$, the direction ($\hat p$) and strength ($A$) of modulation signal present in a given map can be recovered. 

Here we assess the statistical nature of the anisotropic signal underlying an nSI map, by reading the pixel values of $\xi_r (\hat{N})$ in $\hat{p}$ and $-\hat{p}$ directions. For the SID modulation case, we expect to
find $\langle \xi_r (\hat{p}) \rangle = \langle \xi_r (-\hat{p}) \rangle = 2A$.
In the case of an nSI map with scale dependent modulation amplitude, the angular dependence of the intrinsic modulation strength can be profiled by estimating $\xi_r(\hat{N})$ for different choices of disc radii `$r$' and obtaining its dipole component thereof.

\bibliographystyle{jcappub}
\bibliography{ref}
\end{document}